\documentclass{PoS}

\usepackage{color}
\usepackage{epsfig}

\newcommand{\beqn}{\begin{eqnarray}}
\newcommand{\eeqn}{\end{eqnarray}}
\newcommand{\eq}[1]{(\ref{#1})}
\newcommand{\preprintline}{\newline
\vskip -4.2cm
\rightline{\parbox{4cm}{\large\rm LU-ITP 2005/020\\ ITEP-LAT/2005-18\\ HU-EP-05/56}}
\vspace{3.2cm}}

\PoS{PoS(LAT2005)295}

\title{An Abelian two-Higgs model and high temperature superconductivity\preprintline}

\ShortTitle{An Abelian two-Higgs model and high temperature superconductivity}

\author{M.~N.~Chernodub\thanks{Supported by the grants
        RFBR 04-02-16079, RFBR 05-02-16306,
        DFG 436 RUS 113/739/0, MK-4019.2004.2.}\\
        ITEP, Bolshaya~Cheremushkinskaya 25, 117259 Moscow, Russia\\
        E-mail: \email{Maxim.Chernodub@itep.ru}}
\author{\speaker{Arwed~Schiller}\\
        Institut f\"ur Theoretische Physik,
        Universit\"at Leipzig, D-04109 Leipzig, Germany\\
        E-mail: \email{Arwed.Schiller@itp.uni-leipzig.de}}
\author{E.-M.~Ilgenfritz\thanks{Supported by
        the DFG Forschergruppe 465 ``Gitter-Hadronen-Ph\"anomenologie''}\\
        Institut f\"ur Physik, Humboldt-Universit\"at zu Berlin,
        Newtonstr. 15, D-12489 Berlin, Germany\\
        E-mail: \email{ilgenfri@physik.hu-berlin.de}}

\abstract{We study a three dimensional Abelian Higgs model
containing singly- and doubly-charged scalar fields coupled to a
compact Abelian gauge field in the London limit. The model
attracts interest because of its relevance to high-$T_c$
superconductors with charge 1 holon and charge 2 spinon-pair
fields. It contains two types of vortices carrying magnetic flux
and one type of instanton-like monopoles. Using thermodynamic and
topological observables we present the phase diagram in the
parameter space of the gauge and holon and spinon-pair couplings.
The Fermi liquid, the spin gap, the superconductor and the strange
metallic phases have been identified in a wide region of
parameters. The model may serve as a toy system modelling
non-perturbative properties of the Yang-Mills theory.}
\FullConference{XXIIIrd International Symposium on Lattice Field Theory\\
 25-30 July 2005\\
 Trinity College, Dublin, Ireland}

\begin{document}

\section{Introduction}

The physics of high-$T_c$ superconductivity~\cite{ref:highTc} is
not understood yet. At normal temperatures, all known high-$T_c$
superconductors are ceramic crystals characterized by a poor
conductivity. At low temperatures the clean ceramic materials are
rather insulators than conductors (Mott insulators).
However, as one adds impurities
to the clean material (``doping''), a good insulator becomes in
certain cases a superconductor at low enough temperatures.

Despite superconducting specimen are three-dimensional structures
the physics of high-$T_c$ superconductivity is believed to be
essentially two-dimensional~\cite{ref:highTc}. In fact, all known
high-$T_c$ superconductors consist of
copper oxygen (CuO${}_2$) lattice planes.
In the undoped state the crystal does not contain enough free carriers
of electric charge. The role of impurities is to provide
the carriers -- electrons
(like in  $\mbox{Nd}_{2-x}\mbox{Ce}_{x} \mbox{Cu} \mbox{O}_4$ case)
or holes
(like in $\mbox{La}_{2-x} \mbox{Sr}_x \mbox{Cu} \mbox{O}_4$ material) --
into the CuO${}_2$ planes,
eventually making the CuO${}_2$ planes superconducting 2D systems.
Therefore, the physics of high-T${}_c$ superconductivity must be
understood as an essentially 2D phenomenon.

Physically, there are two basic parameters: the temperature $T$ and
the concentration of impurities (doping) $x$. The phase diagram
of a real high-$T_c$ superconductor is the
temperature-concentration plane and
typically contains the following phases:
a metallic, a pseudogap (found in hole doped materials),
an anti-ferromagnetic and a superconducting phase.

One popular approach to the physics of superconducting planes
is based on a simplified model describing the dynamics of holes and spins.
Basically we have a lattice with hopping holes and localized spins acting
as dynamical variables. The lattice spacing and possible anisotropy is
dictated by the ceramic host material itself.
The effective description is provided by the $t-J$
Hamiltonian~\cite{ref:highTc}:
\beqn
  H_{tJ} = - t  \sum\limits_{<ij>,\sigma} c_{i\sigma}^\dagger
  (1 - n_{i,-\sigma}) (1- n_{j,-\sigma})c_{j\sigma}
  + J \sum\limits_{<ij>} ({\vec S}_i {\vec S}_j - \frac{1}{4} n_i n_j)
  \,,
  \label{eq:tJ}
\eeqn
where the first term describes the hopping of holes or electrons without
changing spin while
the second term describes the anti-ferromagnetic Heisenberg coupling
between the spins located at
the copper-sites. Here
$\vec S_i=(1/2) \sum_{\sigma\sigma'}c_{i\sigma}^\dagger {\vec \sigma}_{\sigma\sigma'}  c_{i\sigma'}$
is the spin operator, $c_{i\sigma}$ and $n_{i,\sigma}$ are the electron creation and
occupation number operators for given spin $\sigma$, respectively,
and $n_i=\sum_\sigma c_{i\sigma}^\dagger c_{i\sigma}$.

Even the simplified $t-J$ model~\eq{eq:tJ} is difficult to solve.
One successful approach is based on the slave boson formulation~\cite{ref:slave-boson},
which proposes to split the spin and charge variables of the electrons.
Let the electron creation operators be written
as $c^\dagger_{i\sigma} = f^\dagger_{i\sigma} b_i$, where $f_{i\sigma}$
is a spin-particle
(``spinon'') operator and $b_{i}$ is a charge-particle (``holon'')
operator. In order to forbid double occupancy of sites
one imposes the constraint $f^\dagger_{i\uparrow} f_{i\uparrow} + f^\dagger_{i\downarrow}
f_{i\downarrow} + b^\dagger_{i} b_{i} = 1$ on the physical states
of the system.

The spinon is a fermionic chargeless particle which
carries information about the spin of the electron while the holon
is a bosonic particle which is responsible for the electron charge.
The essential feature of the spin-charge separation approach
is that it introduces an additional (internal) compact $U(1)$ degree
of freedom which can be formulated as a gauge freedom on the operator level:
\beqn
  c_{i\sigma} & \to & e^{i \gamma_i} c_{i\sigma}\ : \quad
  f_{i\sigma} \to f_{i\sigma}\,, \quad
  b_{i} \to e^{i \gamma_i} b_{i\sigma}\quad\ \mbox{[usual gauge freedom]}
  \,,
  \label{eq:gauge:external}\\
  c_{i\sigma} & \to & c_{i\sigma}\ :\quad
  f_{i\sigma} \to e^{i \alpha_i} f_{i\sigma}\,, \quad
  b_{i} \to e^{i \alpha_i} b_{i\sigma}\quad \mbox{[internal gauge freedom]}
  \,.
  \label{eq:gauge:internal}
\eeqn

\newpage

\section{The compact $U(1)$ gauge model coupled to matter}
\label{sec:compactU1}

The emerging effective theory of superconductivity can further be
simplified and reformulated as a lattice gauge
model~\cite{ref:LeeNagaosa:characterization}. The basic idea is to
neglect -- or absorb into the couplings of the effective model --
the external gauge degree of freedom~\eq{eq:gauge:external} and to
concentrate attention to the internal compact gauge degree of
freedom~\eq{eq:gauge:internal}. In fact, the coupling of the usual
electromagnetic field to the charge-carrying holon degrees of freedom
(as well as the inter-holon interaction due to the electromagnetic
interaction) is relatively weak compared to the strong correlations
among the electrons in the lattice environment.

Thus, we go over from the $t-J$ model~\eq{eq:tJ} to a compact Abelian
gauge model with internal symmetry~\eq{eq:gauge:internal}, which
couples holons and spinons.
As in usual BCS superconductivity, at certain parameters
of the $t-J$ model the spinons couple and form
bosonic quasiparticles. In a mean field theory one can define the fields
\beqn
  \chi_{ij}
  = \sum_\sigma \langle f^\dagger_{i\sigma} f_{j\sigma} \rangle
  \to \chi_{ij} \cdot
  {\mathrm e}^{- i (\alpha_i -\alpha_j)}\,,
  \quad
  \Delta_{ij} = \langle  f_{i\uparrow}f_{j\downarrow}
                       - f_{i\downarrow}f_{i\uparrow}  \rangle
  \to \Delta_{ij} \cdot
  {\mathrm e}^{i (\alpha_i + \alpha_j)}\,,
\eeqn
which behave under the internal gauge transformations~\eq{eq:gauge:internal}
as a neutral (vector-like) particle and a doubly-charged matter field,
respectively. The phase of the $\chi$-particle is associated with the compact
$U(1)$ gauge field, $\theta_{ij} \equiv \arg \chi_{ij} \to
\theta_{ij} - \alpha_i + \alpha_j$. The radial part of the $\chi$-particle
defines the so-called ``resonating valence bond'' (RVB) coupling
$\chi = \langle |\chi_{ij}| \rangle$,
and the doubly-charged matter
field $\Delta$ is called spinon-pair field (an analog of the doubly-charged
Cooper pairs).

Usually the RVB coupling is treated in the mean-field approximation
and therefore is assumed to be fixed, $\chi(x) = \chi \equiv {\mathrm{const}}$.
Thus, the dynamical content of the effective model is given by
a singly-charged boson field (holon) $b$, a doubly-charged boson field (spinon-pair)
$\Delta$ and the compact $U(1)$ gauge field $\theta$.

\begin{figure}[!htb]
  \centerline{\includegraphics[scale=0.45,clip=true]{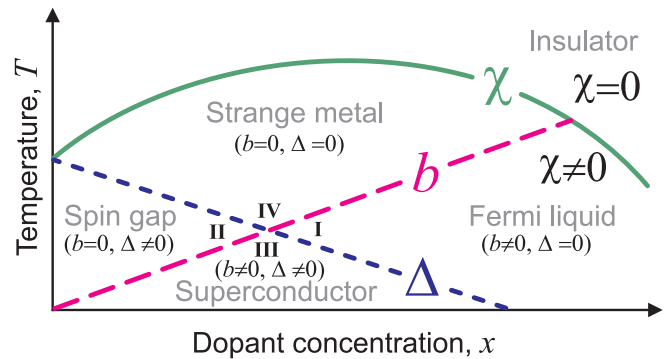}}
  \caption{The schematic phase diagram of the $U(1)$ model for a high-$T_c$
           superconductor~\cite{ref:LeeNagaosa:characterization}.
          }
  \label{fig:lee:nagaosa:phase}
\end{figure}

At high temperature the RVB coupling is vanishing, $\chi=0$, and
the system is in the Mott insulator (or, poor metallic) phase,
see Fig.~\ref{fig:lee:nagaosa:phase}.
As the temperature decreases,
the RVB coupling is getting non-zero, $\chi \neq 0$,
enabling the formation of the spinon-pair condensate
$\Delta = |\langle\Delta_{ij}\rangle|$
and/or of the holon condensate
$b = \langle b_i\rangle$.
Depending on the presence of the condensates, four phases
classified in Ref.~\cite{ref:LeeNagaosa:characterization}
may emerge. In Fig.~\ref{fig:lee:nagaosa:phase}
they are sketched and denoted as the Fermi liquid, the spin gap,
the superconductor, and the strange metallic phases.

\newpage

\section{The compact Abelian two-Higgs model on the lattice}

Following the proposal in Ref.~\cite{ref:LeeNagaosa:characterization}
the model described in Section~\ref{sec:compactU1} can be studied as compact
Abelian two-Higgs model (cA2HM) in {\it three} dimensions with a $U(1)$
gauge link field $\theta_l$, a single-charged holon field $\Phi_1$,
and a double-charged spinon-pair field $\Phi_2$.
The qualitative characteristics of the model become already visible
in the London limit, in which the radial parts of both
Higgs fields $\Phi_i = |\Phi_i| \, e^{i \varphi_i}$ are frozen,
$|\Phi_i| = {\mathrm{const}}$, $i=1,2$.
The remaining dynamical Higgs variables are the
phases of the matter fields $\varphi_{1,2}$. The action of the cA2HM model
in the London limit is:
\beqn
  S[\theta,\varphi_1,\varphi_2] = -\beta \sum_P \cos\theta_P
           - \kappa_1 \sum_l \cos({\mathrm{d}} \varphi_1 + \theta)_l
           - \kappa_2 \sum_l \cos({\mathrm{d}} \varphi_2 + 2 \theta)_l\,,
  \label{eq:cA2HM:action}
\eeqn
where $\theta_P$ is the standard lattice plaquette field, ${\mathrm{d}}$
denotes the ordinary lattice derivative,
${({\mathrm{d}} \varphi)}_{x,\mu} = \varphi_{x+\hat\mu} - \varphi_{x}$.
The parameter $\beta$ is the inverse gauge coupling,
$\kappa_1 \propto t \cdot x$ and $\kappa_2 \propto J$ are the hopping
parameters of the holon and the spinon-pair field, respectively.
They are proportional to the parameters of $t-J$ model.
The model~\eq{eq:cA2HM:action} obeys the $U(1)$ gauge invariance:
\beqn
  \theta \to \theta + {\mathrm{d}} \alpha\,,\qquad \varphi_1
  \to \varphi_1 -  \alpha\,, \qquad
  \varphi_2 \to \varphi_2 -  2 \alpha\,,
\eeqn
which is a direct counterpart of the internal gauge
symmetry~\eq{eq:gauge:internal}.

In the limit $\kappa_2 \to 0$ one gets the compact
$U(1)$ gauge model with a charge-1 Higgs field only, the holon.
The phase diagram of the $Q=1$ compact Abelian Higgs model
contains~\cite{ref:fradkin,ref:AHM1}
two regions: the confinement ``phase'' (characterized by a low holon condensate
$b$) at small hopping parameter $\kappa_1$) and the Higgs ``phase'' at large
$\kappa_1$ (where the condensate $b$ becomes large). Both regions are
analytically connected
for strong enough Higgs selfcoupling, although
a Kert\'esz line~\cite{ref:Arwed} {\it physically} separates them also there.
A {\it qualitatively} similar picture (however with a true
second order phase transition)
is realized~\cite{ref:AHM2} in the limit $\kappa_1 \to 0$
where one gets the $Q=2$ compact Abelian Higgs model in which
the spinon-pair field plays the role of the sole Higgs field.
Thus, we expect that the phase diagram of the
model~\eq{eq:cA2HM:action} should contain all four phases depicted in
Fig.~\ref{fig:lee:nagaosa:phase} in the $\chi>0$ region.

Below we report our Monte Carlo investigation of the cA2HM which is still
under way.
To scan the phase diagram we have simulated on a $16^3$ lattice, choosing
three values of the gauge couplings, $\beta=1.0$, $1.5$ and $2.0$,
for a huge grid of $\kappa_{1,2}$ hopping parameter pairs covering
the range\footnote{The actually
relevant region of the $\kappa_1$-$\kappa_2$ plane depends on $\beta$.}
$0 < \kappa_{1,2} \leqslant 2.5$.
We are particularly interested in the strongly coupled case, $\beta \simeq 1$.

The compactness of the gauge field guarantees the presence of
instanton--like magnetic mo\-no\-po\-les. An elementary monopole is
a source of $F^{\mathrm{mon}} = 2 \pi$ units of magnetic flux
associated with the internal gauge field $\theta$.
Due to the matter fields two types of topologically stable vortices exist.
The magnetic flux quanta of the vortices corresponding to the
holon and the spinon-pair ``Higgs'' fields are
$F^{\mathrm{vort}}_1 = 2 \pi$ and $F^{\mathrm{vort}}_2 = \pi$,
respectively. Since the magnetic flux is conserved, one
monopole is simultaneously a source of one holon vortex and
two spi\-non-pair vortices.

\begin{figure}[!htb]
  \begin{tabular}{ccc}
     \includegraphics[scale=0.31,clip=true]{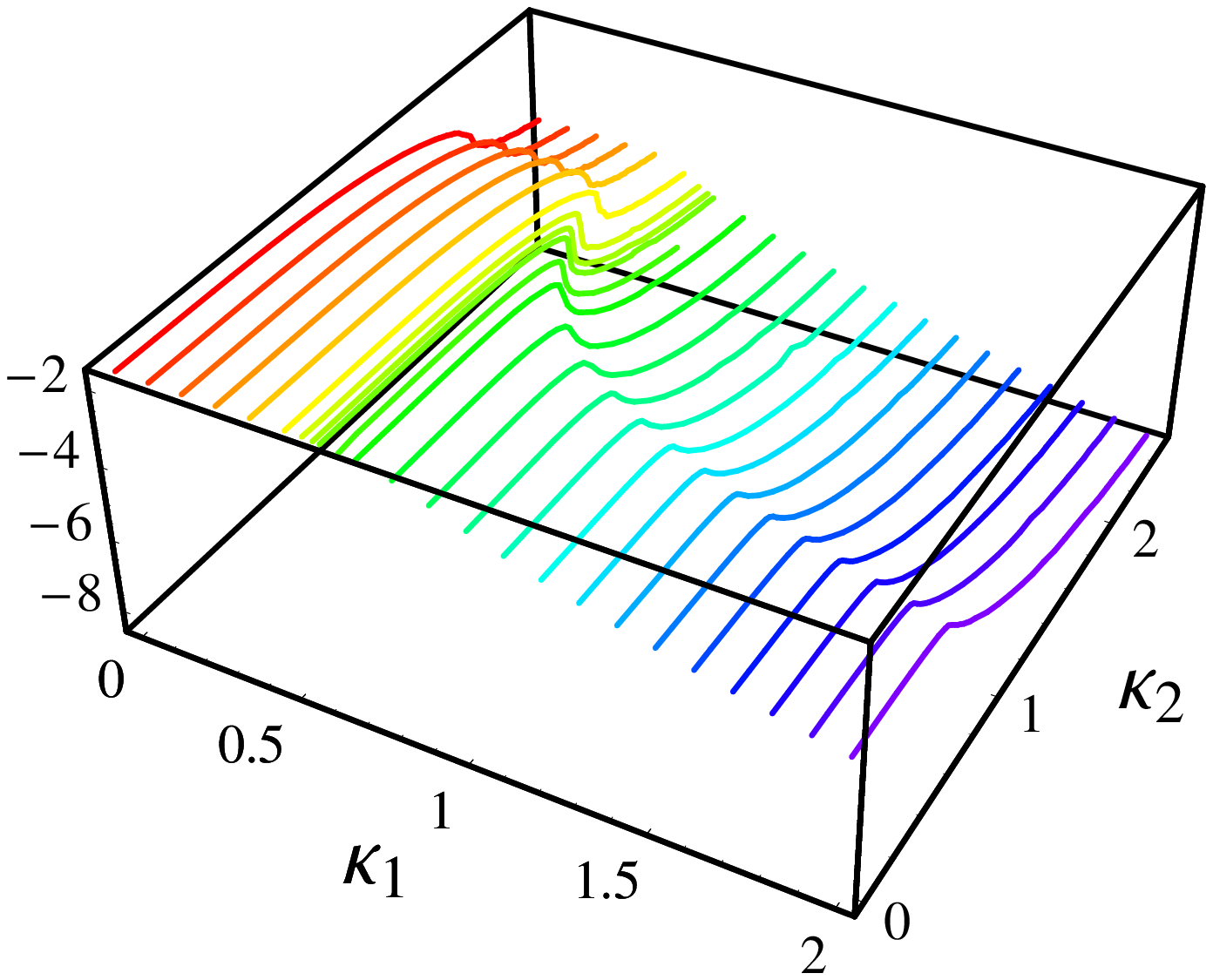} &
     \includegraphics[scale=0.31,clip=true]{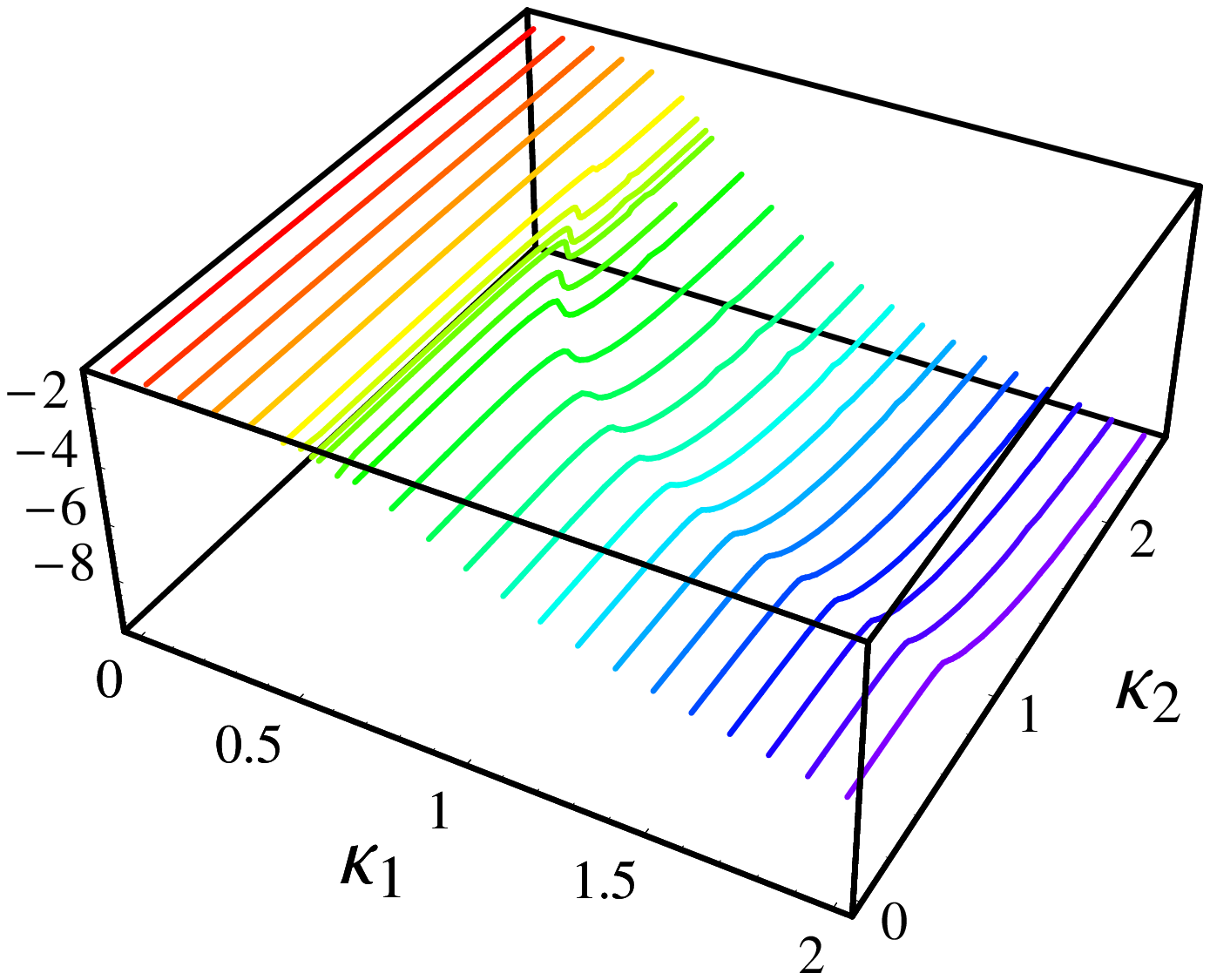} &
     \includegraphics[scale=0.31,clip=true]{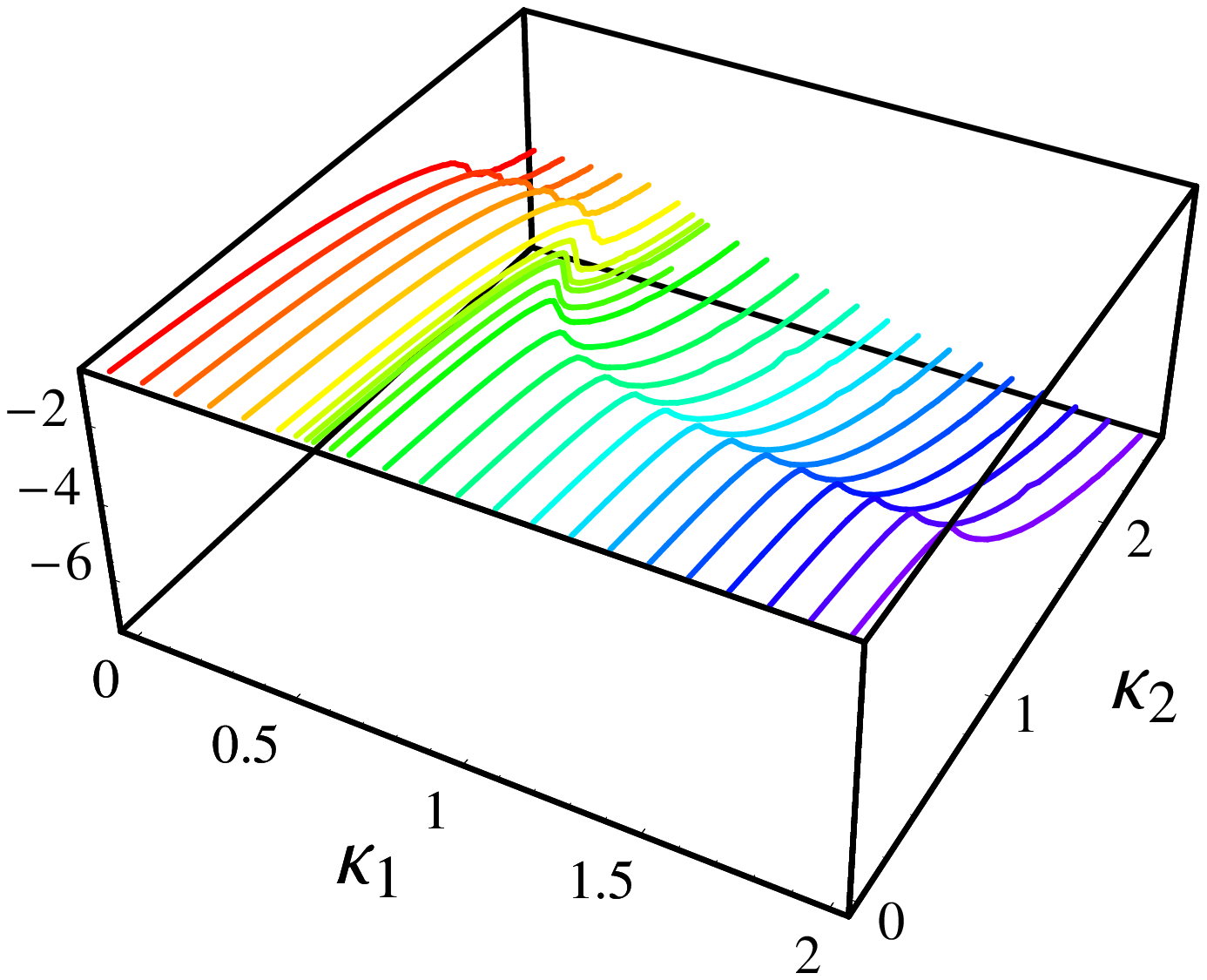} \\
     monopole & holon vortex & spinon-pair vortex\\
  \end{tabular}
  \caption{The densities of the topological defects over the
           $(\kappa_1,\kappa_2)$ plane at $\beta=1$.}
  \label{fig:density}
\end{figure}

The densities of the three topological defects are logarithmically
plotted in Fig.~\ref{fig:density} over the $(\kappa_1,\kappa_2)$ plane
for a strong gauge coupling $\beta=1$.
With increasing hopping parameters the monopole density gets suppressed.
The density of the holon (spinon-pair) vortex becomes suppressed
with increasing holon (spinon-pair) hopping parameter $\kappa_1$ ($\kappa_2$),
beyond a line in the $\kappa_1$-$\kappa_2$ plane.

\begin{figure}[!htb]
  \begin{tabular}{ccc}
    \includegraphics[scale=0.31,clip=true]{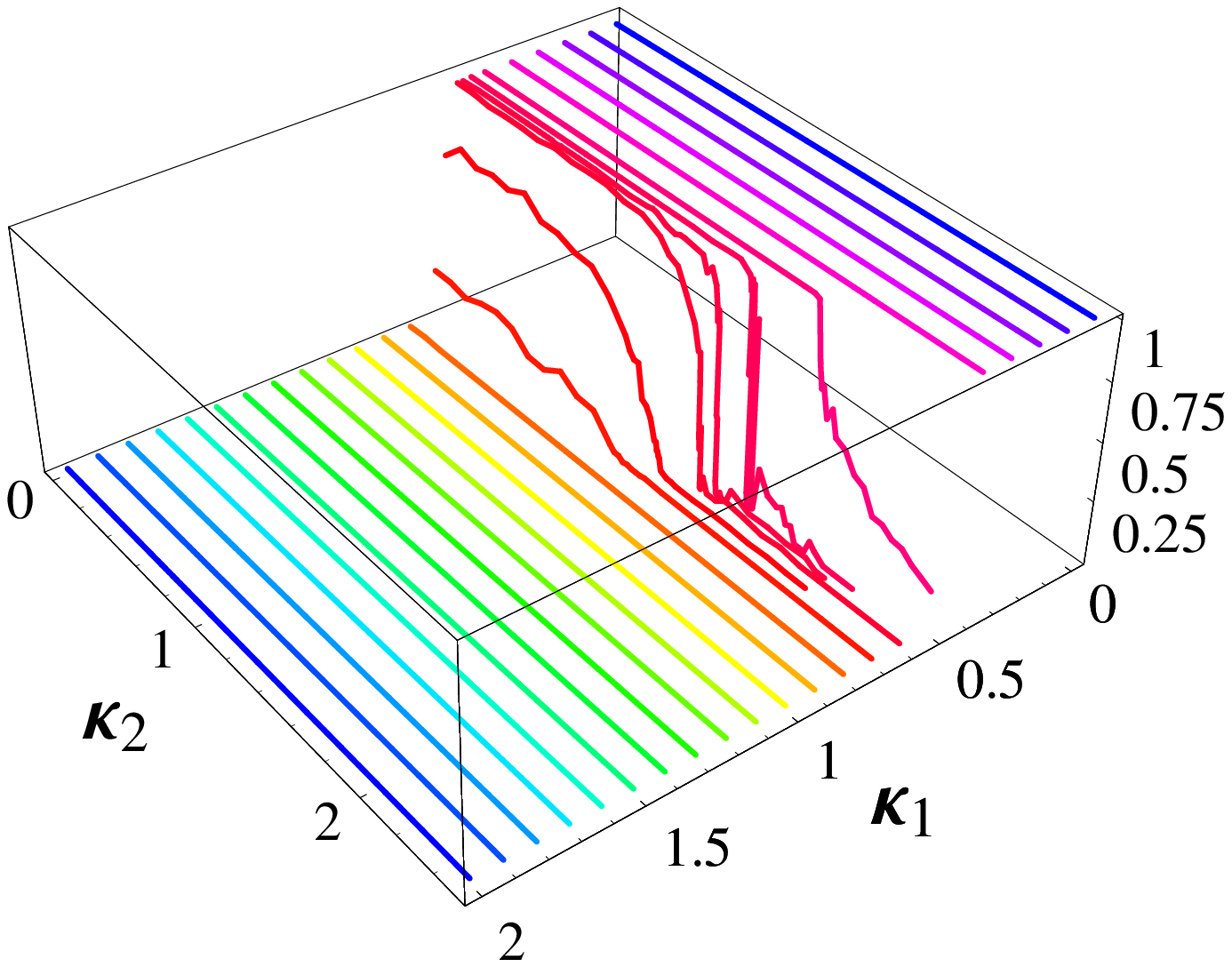} &
    \includegraphics[scale=0.31,clip=true]{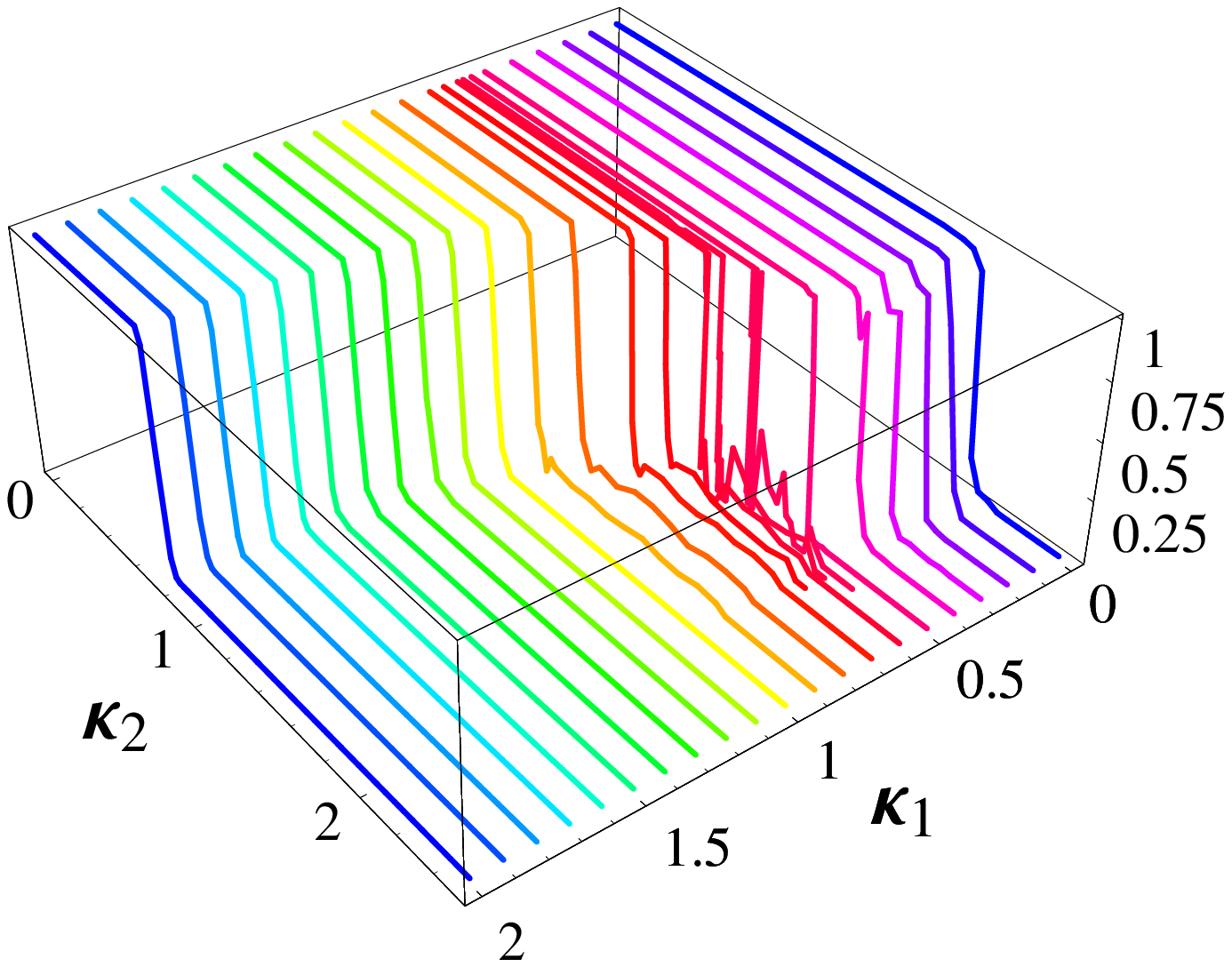} &
    \includegraphics[scale=0.18,clip=true]{beta10crit_proc.eps} \\
    holon vortex& spinon-pair vortex & phase diagram \\
  \end{tabular}
  \caption{The percolation probability of the infrared vortices and the
           phase diagram of the system at $\beta=1$.}
  \label{fig:density:IR}
\end{figure}

\begin{figure}[!htb]
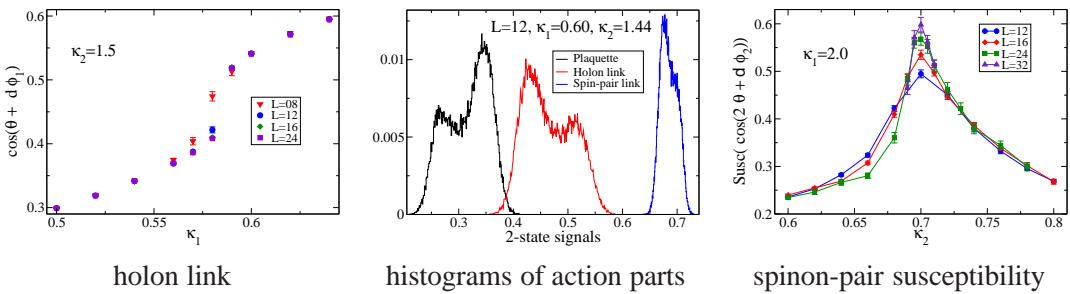

  \begin{tabular}{ccc}
    \includegraphics[scale=0.18,clip=true]{theta_q1_b100_proc.eps} &
    \includegraphics[scale=0.18,clip=true]{hist_proc.eps} &
    \includegraphics[scale=0.18,clip=true]{susc_theta_q2_b100_k1_200.eps} \\
    holon link& histograms of action parts&  spinon-pair susceptibility\\
  \end{tabular}
  \caption{Examples of thermodynamic behavior at various $\kappa_1$,
           $\kappa_2$ and $\beta=1$.}
  \label{fig:thermodyn}
\end{figure}

The connectivity of the vortex clusters gives a clear view of the phase diagram of the model~\cite{ref:AHM1,ref:AHM2,ref:Arwed}.
A Higgs condensate suppresses the proliferation of the corresponding
vortices: they are prevented to percolate over infinitely long distances.
We show in Fig.~\ref{fig:density:IR}
the percolation probabilities dropping to zero (extracted
from the cluster correlation functions) and the corresponding phase
diagram (classified according to Ref.~\cite{ref:LeeNagaosa:characterization},
see Fig.~\ref{fig:lee:nagaosa:phase}).
To identify the phases, we measured the average plaquette
and both link contributions of the action and their susceptibilities~(\ref{eq:cA2HM:action}), respectively.
Some examples are shown in Fig~\ref{fig:thermodyn}.
From a preliminary finite size analysis from $12^3$ to $32^3$ we found:
i) Hint for a first order transition
between III and IV in the crossing region of the two percolation lines
at strong gauge coupling,
ii) No transition along the red (vertical) line for small $\kappa_2$ in agreement
with the limit $\kappa_2\to 0$, \
iii) Signals for thermodynamic transitions along the remaining transitions
lines (second order along the horizontal parts of the blue line).

\section{Summary}

We have observed two transitions associated with the patterns of
vortex percolation in the Abelian two--Higgs model of high-$T_c$
superconductivity.
These transition lines are roughly parallel to the corresponding
hopping parameter axes $\kappa_i$, almost perfectly parallel
at weak gauge coupling ($\beta=2$).
The crossing in some region of finite ($\kappa_1,\kappa_2$)
gets new features at strong coupling $\beta=1$.
The Fermi liquid, the spin gap, the superconductor
and the strange metallic phases
have been identified in a wide region of parameters of this model.
The percolation transitions are accompanied with ordinary
phase transitions except at small $\kappa_1$ below the crossing
region of the two transition lines.
First hints for a changing order
along the thermodynamic transition lines are found, and the
joint transition in the crossing region seems to be first order.
Further studies are underway to fortify our findings.

More complex Higgs sector will be necessary to build more
realistic effective models reproducing the confinement mechanism
of gluodynamics in the sense of Ref.~\cite{ref:AHM2}. The
spin-charge separation idea of the high-$T_c$ superconductivity
should further be extended to gluodynamics to reformulate it in
terms of condensed matter systems~\cite{ref:nematic:hightc}.

\end{document}